\documentclass[newversion]{aa}
\usepackage{graphicx}

\begin{document}


   \title{The Deep Layers of Sunspot Umbrae}

   \author{G. Stellmacher\inst{1}
          \and E. Wiehr\inst{2}}

   \offprints{E. Wiehr}

   \mail{ewiehr@astrophysik.uni-goettingen.de}

   \institute{Institute d'Astrophysique (IAP), 98 bis Blvd. d'Arago, 
              75014 Paris, France
Institut f\" ur Astrophysik der Universit\"at,
              Friedrich-Hund-Platz 1, 37077 G\"ottingen, Germany
              \and
              Institut f\" ur Astrophysik der Universit\"at,
              Friedrich-Hund-Platz 1, 37077 G\"ottingen, Germany}

   \date{Received Jan. 14, 1975; accepted Aug. 8, 1975}

\abstract
{}
{We model the deepest observable layers of dark sunspot umbral atmospheres in terms
of an empirical model which equally describes observed near infrared continuum 
intensities and line profiles.}
{We use the umbral continuum intensity at $1.67\mu$ and the three C\,I lines at 
16888\AA{}, 17449\AA{} and 17456\AA{} to model the deep layers near the minimum of 
H$^-$ absorption. An extrapolation of umbra models to such deep layers must assure 
that (a) the calculated 1.67$\mu$ continuum does not fall below the range of 
observations and (b) the resulting C\,I lines do not come out stronger than the 
lower observational limit.}
{Our calculations show that a T$^4(\tau_R)$ stratification yields the best 
compromise between both criteria: a flatter gradient violates (a) a steeper one 
(b). We determine $T_{eff}$ from the umbral and photospheric flux ratio down-scaling 
the monochromatic photospheric flux with the umbral contrast for each frequency 
$\nu$ and obtain the monocharomatic umbral flux. Integration of both over 
$\nu$ gives the ratio of total umbral and photospheric flux, which yields 
$3560<T_{eff}<3780$\,K. We assume for our model $T_{eff}=3750$\,K and fit it to 
the theoretical model by Meyer et al. (1974). Comparison of the gradient 
$\nabla =(d\,logT)/(d\,logP_g)$ with the adiabatic one shows that umbral 
convection, if existing at all, can only occur at considerably deeper layers 
than in the photosphere.} 
{}
\keywords{Sunspot umbra - infrared contrasts - line profiles - empirical model - 
effective temperature}

\maketitle

%
%

\section{Introduction}

Recent observations of umbral continuum contrasts in the near infrared, carried out by Hall 
(1970) and by Ekmann and Maltby (1974), have led to a re-discussion of the temperature and 
pressure stratification in sunspot umbrae (Zwaan, 1974; Kjeldseth-Moe and Maltby, 1974). 
Their models disagree considerably with the empirical model M0 by Stellmacher and Wiehr 
(1970, hereafter referred to as Paper\,I), which was shown to optimally fit the profiles 
of several magnetically insensitive lines observed in umbrae and had been confirmed by 
Kneer (1972) and by Koppen (1974). In Stellmacher and Wiehr (1971, hereafter referred 
to as Paper\,II) this fit was extended to the center-to-limb variation of a non-split 
line, which represents a very strong criterion for empirical models.

M0 had been deduced from line profiles and continuum contrasts at 
5000\AA{}$<\lambda<8000$\AA{}, and can thus not represent the deep layers 
$\tau_{0.5}\ge1.62$, for which it gives an arbitrary extrapolation. The present study 
demonstrates how M0 can be extended towards deeper layers to fit observed IR data. 
The still existing scatter of observational data and of the photospheric models 
(to which any umbra model necessarily refers) allows an extention of MO to deep 
layers preserving the good line profile presentation in papers\,I and II without
assumption of additional parameters. We consider our empirical model as a 'boundary 
condition for three dimensional sunspot models' and for the 'quantitative 
interpretation of line profiles' (Zwaan 1974).

\section{Observational data}

Considering the 'epochal decrease' of observed umbra-to-photosphere intensities in 
the visible region from 0.24 (at 5452\,\AA{}, Michard 1953) to 0.04 (at 5790\,\AA{}, 
Ekmann and Maltby 1974), recently observed infrared data may hardly be considered as 
'final results'. In particular, the determination of an empirical umbra model from 
the low data in the visible range together with recent high infrared data will be 
uncorrect. Besides, differences between individual umbrae require the use of 
{\it data from one spot taken under identical conditions} such as the simultaneous 
broad band measurements by Ekmann and Maltby (1974).
                    
%
   \begin{table}[h]
      \caption[]{Correction factors for absorption lines in 
the broad-band wavelength regions used by Ekmann \& Maltby (1974).}
   \hspace{20mm}\includegraphics[width=5.0cm]{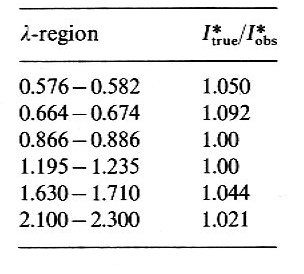}
   \label{tab1}   
   \end{table}
%
%

We correct their data for the influence of Fraunhofer lines integrating the ratio 
$I^{umbra}/I^{phot}$ from Hall's (1974) infrared atlas in the respective $\lambda$-windows. 
For the visible region the same is done for the data by W\"ohl (1975). The correction 
factors, given in Table\,1, might still be underestimated due to the uncertain 
continuum level.
 
%

   \begin{figure}[ht]     
   \hspace{-4mm}\includegraphics[width=9.8cm]{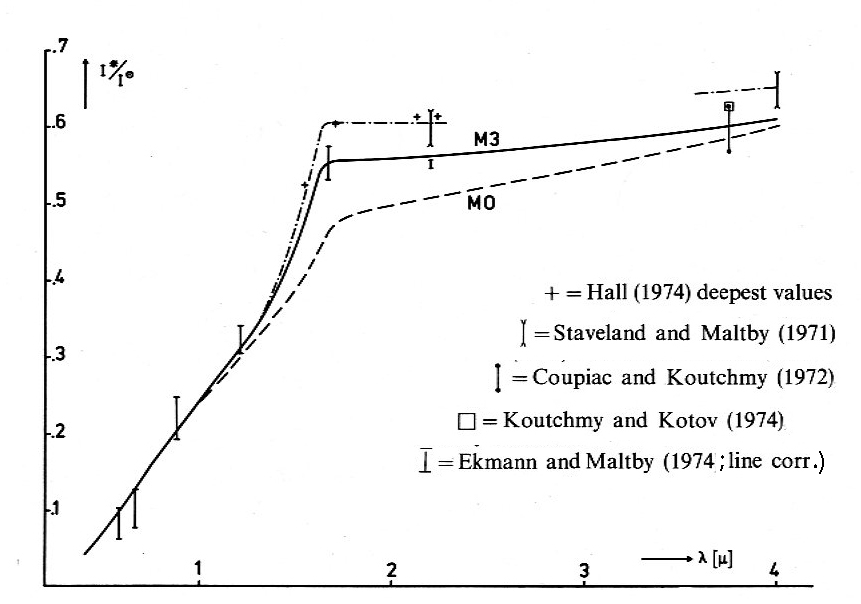}
   \caption{Observed relative umbral continuum intensities compared to calculations with our
former model M0 ({\it dashed line}) and with the present model M3 ({\it solid line}, both 
relative to Labs \& Neckels 1968); {\it dash points} = M3 relative to Peteyraux (1952) and 
Badinov et al. (1965). Bars  give raw and maximum corrected observations by various observers.}
   \label{Fig1}
    \end{figure}

We finally applied these corrections to the data from all umbrae at $\vartheta \le 20^o$
given by Ekmann and Maltby (1974). The results are given in Fig.\,1 together the umbral 
intensities by Hall (1974), which appear to be considerably higher. Hall's data refer to 
continuum windows and are thus not affected by Fraunhofer lines; only parasitic light 
may interfere which does not exceed 2\% considering the Fe\,II lines in that 
$\lambda$-region to be of non-umbral origin (cf., paper\,I). The remaining differences 
to the Ekmann and Maltby (1974) data cannot be analyzed, since Hall (1974) did not observe 
in the visible $\lambda$-range; light bridges in his spots may also explain his higher 
intensity values.

\section{Extrapolation of M0 to the improved M3 model}   
    
It has been shown by Zwaan (1974) that the strong variation of relative umbra intensity 
between the maximum $H^-$ absorption at $0.8\,\mu$ and the absorption minimum at 
$1.6\,\mu$ requires the steepest possible temperature gradient, i.e. radiative equilibrium 
whith $T^4=3/4\cdot T_{eff}^4(\tau_R + q)$, where $T_{eff}$ and $\tau_R$ denote the effective 
temperature and the Rosseland opacity, respectively and $q$ is constant for deep layers. 
We therefore assume a $T^4(\tau_R)$ gradient for the extrapolation of our model M0 to deep layers. 

To estimate the actual umbral $T_{eff}$ value we multiplied the monochromatic photospheric 
flux $F_{\nu}^{phot}$ (Uns\"old, 1956, Fig.\,20a) by the observed monochromatic intensity 
ratio $I^{umbra}_{\nu}/I^{phot}_{\nu}$. This yields the monochromatic umbral flux 
$F_{\nu}^{umbra}$ under the assumption of a vanishing center-to-limb variation of the 
contrast at all $\nu$, which is found in observations and in all umbral models under 
discussion (cf., paper\,II). Planimetry over the frequency $\nu$ then yields for umbra 
and photosphere their total (integrated) flux $F^{umbra}$ and $F^{phot}$.

Assuming now radiative equilibrium, r.e., it is $F^{umbra}/F^{phot}=T^4_{eff}(umbra)/T^4_{eff}(phot)$,
and we obtain 3560\,K$<T_{eff}^{umbra}<3780$\,K; the scatter reflects the range of observed contrasts. 
In order to keep the temperature gradient as high as possible we assume $T_{eff}^{umbra}=3750$\,K; the 
resulting model M3 is listed in Table\,3, it is shown in Fig.\,2 together with M0 and some other models.
The validity of this M3 has now to be checked by comparison with line profile and continuum observations.
                      
%
   \begin{figure}[hb]     
   \hspace{-1mm}\includegraphics[width=8.9cm]{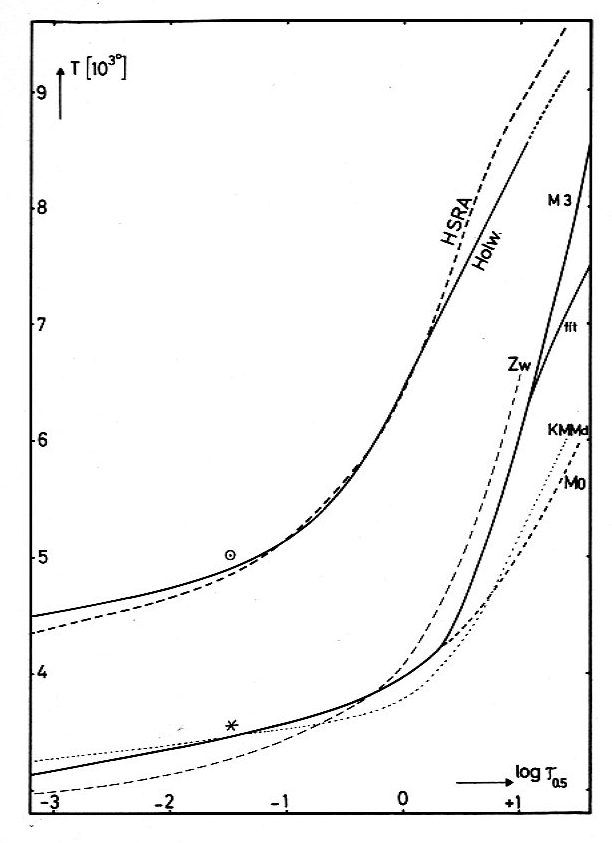}
   \caption{Umbral models: Zw=Zwaan (1974), KMMd=Kjeldseth-Moe \& Maltby (1974) dark, M0 our 
former, M3 the present model, 'fit' = adaption to Meyer et al. (1974); compared to photosphere 
models: Holw=Holweger (667), HSRA=Gingerich et al. (1971).}
   \label{Fig2}
    \end{figure}

%
%

\subsection{The M3 representation of the continuum}

In order to avoid uncertainties from the deep layers of photospheric models {\it we normalize 
our calculated umbra intensities exclusively to observed photospheric absolute intensities}: 
Labs and Neckel (1968) list for $1.24\mu\le\lambda \le3.6\mu$ the relative measurements by Pierce 
(1954), which they calibrated using the 'model distribution between $1.0\,\mu$ and $1.5\,\mu$ as a 
linking medium'. This affects that the two intensity values given by Pierce for $0.99\,\mu$ and for 
$1.05\,\mu$ are $6\%$ too high. If we accordingly correct the relative intensities by Pierce (1952), 
we obtain a remarkable agreement with Peyturaux (1952).

A similar uncertainty concerns the photospheric absolute intensities at $3.6\mu\le\lambda\le4.8\mu$ 
which, according to Koutchmy and Peyturaux (1970) show a large scatter among the various observations. 
In particular, Badinov et al. (1965) publish absolute intensities, which are considerably lower 
than those listed by Labs and Neckel (1968). 

Consequently we normalize for those two wavelength regions the intensities from our umbra model 
alternately to both, the data listed by Labs and Neckel (1965) and the lower values given by 
Peyturaux (1952) and by Badinov et al. (1965), respectively. The resulting relative umbral 
intensities for M0 and for M3, given in Fig.\,1 together with the observations discussed 
in Sec.\,2, show that model M3 fully agrees with the range of continuum observations when 
considering those uncertainties in the photospheric data.

\subsection{The M3 representation of line profiles}

Since model M3 differs from M0 only in the deeper layers $\tau_{o.5}\ge1.6$ (see Fig.\,2 and 
Table\,3) one would not expect significant influence on the line profiles mentioned in Sec.\,1. 
However, calculations show that the contribution of these deeper layers to the line wings of 
NaD$_2$ and of Fe\,5434 is not sufficiently small to be neglected. Moreover, the line profile 
calculations show that M3 represents an 'upper limit' for an extrapolation of M0 to deep layers.

For our line profile calculations we adapted the damping constants $\gamma/\gamma^{theor}$ to 
the observed photospheric line wings taking the known $(g\cdot f)$ values by Lambert and Warner 
(1968) for NaD$_2$ and by Wolnik et al. (1970) for Fe\,5434 together with the relative abundances 
$log(\epsilon_{Na}/\epsilon_H)=-5.75$ and $log(\epsilon_{Fe}/\epsilon_H)=-4.7$, thus improving our 
fit of $log(g\,f\,\epsilon)$ to $W_\lambda^{phot}$ in papers\,I and II. With 
$\gamma_{NaD2}=1.4\,\gamma^{theor}$ and $\gamma_{Fe 5434}= 1.8\,\gamma^{theor}$ we achieve an
agreement within $1\%$ with the observed {\it photospheric} line wings at 
$\Delta \lambda_{Fe 5434}\ge0.1$\,\AA{} and at $\Delta \lambda_{NaD2}\ge0.3$\,\AA{}, 
when using Holweger's (1967) photospheric model and assuming LTE. 
 
The validity of LTE is reasonably be also assumed for the umbral model. The line profiles obtained 
with M3 are shown in Figs.\,7 and 8 together with our observations discussed in Papers\,I and II. 
The bars give the raw data as upper and the maximum uncorrected ones as lower end. The maximum 
correction corresponds to actually vanishing Fe\,II\,6149 and Fe\,II\,7224 lines in the dark 
umbral core. Figs.\,7 and 8 show that the models by Zwaan (1974) and by Kjeldseth-Moe and 
Maltby (1974) yield profiles outside that range.

As a further criterion for the temperature gradient in deeper umbral layers we use the equivalent 
widths of C\,I\,16888\AA{}, C\,I\,17456\AA{} and C\,I\,17449\AA{} which we determine from 
Hall's (1974) infrared atlas to 7.5, 13.0, and 20.0\,m\AA{}. These values represent upper limits 
since any parasitic light would have strengthened them. The quantity ($g\,f\,\epsilon$) was 
determined for the C\,I lines by a fit to the photospheric equivalent widths from Hall's atlas:
$W_{\lambda}^{phot}(\rm C\,16888)=94$\,m\AA{}, $W_{\lambda}^{phot}(\rm C\,17456)=134$\,m\AA{} and 
$W_{\lambda}^{phot}(\rm C\,17449)=168$\,m\AA{} (see Table\,2).

For the calculation of the C\,I lines one has to consider possible influences of the formation 
of CO on the number density of atomic carbon. That of CO depends slightly on the formation of 
OH and H2; the influence of other molecules is negligible. For M3 we find $n_{CO}\approx n_C$ 
at the $\tau_{O.5}\approx5.0$ layer, which also gives the maximum contribution to the C\,I 
lines and their adjacent continuum. The eventual influence of CO-formation on the C-lines 
amounts to $25\%$. The resulting $W_{\lambda}^{umbra}$ (Table 2) indicate that {\it model M3 
represents the 'steepest possible extrapolation' of MO} and gives the best fit to the observations.

%

   \begin{figure}[ht]     
   \hspace{-3mm}\includegraphics[width=9.2cm]{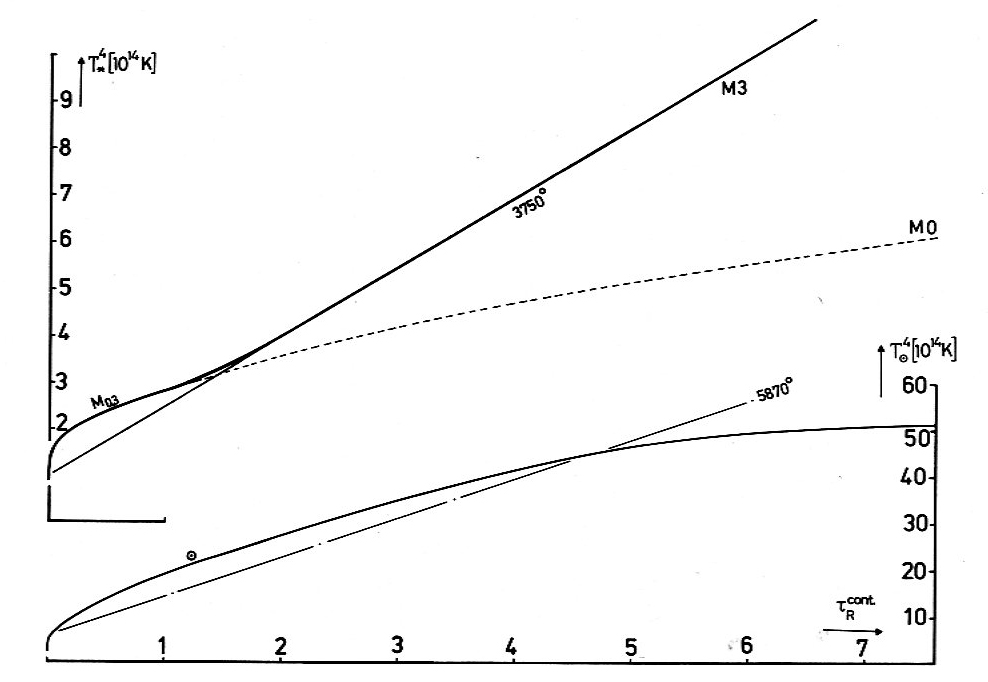}
   \caption{Line blanketing and deviations from rad.equ. for the present model M3 
and the photospheric by Holweger (1967).}
   \label{Fig3}
    \end{figure}

%
%
                    
%

   \begin{figure}[ht]     
   \hspace{-3mm}\includegraphics[width=9.5cm]{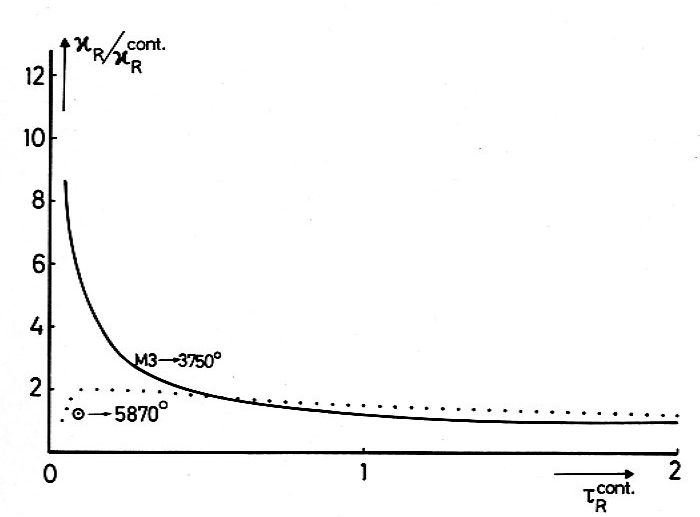}
   \caption{Depth dependence of Rosseland opacity ratio required to convert 
the blanketed into the unblanketed models.}
   \label{Fig4}
    \end{figure}

%
%
                    
%

   \begin{figure}[ht]     
   \hspace{0mm}\includegraphics[width=8.8cm]{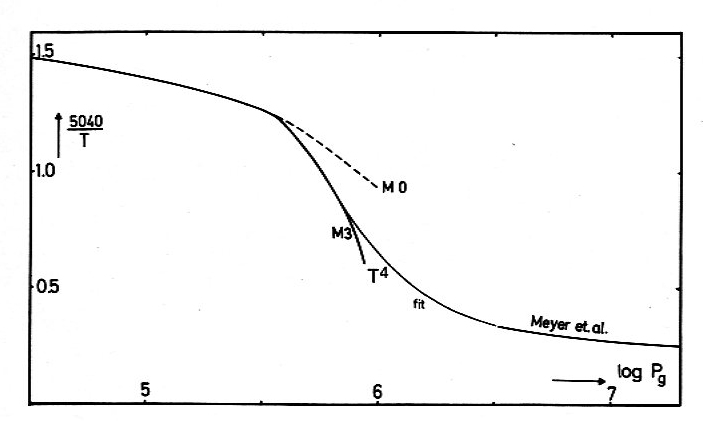}
   \caption{Fit of model M3 to the theoretical umbra stratification by
Meyer et al (1974) without affecting the $1.67\mu$ intensities.}
   \label{Fig5}
    \end{figure}

%
%

%
%
\eject
\section{Conclusion}

Having shown that M3 well represents observations of line profiles and continuum, it 
is reasonable to discuss that model in more detail, keeping in mind that the validity 
of M3 is limited by the accuracy of the observations the model is based on.  

In order to obtain some idea about the {\it umbral line blanketing} we compare M3 with the 
corresponding unblanketed gray r.e. stratification with $T_{eff}= 3750^o$\,K (cf., end of 
Sec.\,3). A presentation in the $T^4(\tau^{contin}_R)$ plane (Fig. 4) shows that M3 has a 
typical line blanketing with 'back-warming' at $1.65\ge\tau_R\ge10^{-3}$ (corresponding to 
$ 2.3\ge\tau_{0.5}\ge3\cdot 10^{-3})$ and 'lowering of the boundary temperature' at 
$\tau_R\le 10^{-3}$, as is predicted by the picket-fence model (Chandrasekhar, 1935).

A more quantitative investigation of the line blanketing can be achieved by transforming the 
blanketed into the unblanketed model. The resulting quantity $\kappa_R/\kappa_R^{cont}$ (Fig. 5) 
shows for M3 a much stronger depth dependence than for the photosphere. Such a behavior has 
already been discussed by Mattig and Schr\"oter (1974).

An interesting question is the {\it limit of radiative equilibrium} (r.e.) below which 
convection is possible. For a rough estimate we connect M3 to the theoretical model by 
Meyer et al. (1974) in the $\Theta=5040/T$ versus $logP_g$ plane (Fig. 6). This is done 
keeping the $1.67\mu$ intensities unaffected within $10^{-3}$ and requires a validity of 
r.e. up to $\tau_{0.5}\le8$. The existing uncertainties in the observed photospheric 
absolute and umbral relative intensities do not allow us to exclude an onset of 
deviations from r.e. at slightly higher layers, which might then give rise to observable 
umbral convection near $1.67\mu$. 
%

   \begin{figure}[h]     
   \hspace{3mm}\includegraphics[width=8.0cm]{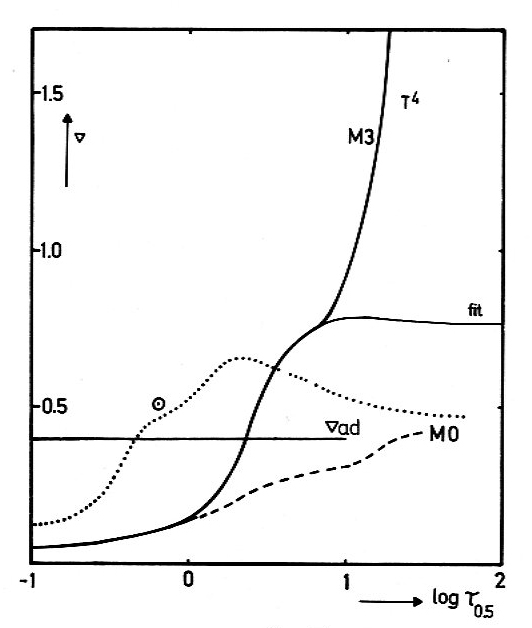}
   \caption{Depth dependence $\nabla$ indicating the much deeper
onset of convection in the umbra as in the photosphere.}
   \label{Fig6}
    \end{figure}

%
%

Another criterion for convection is the maximum in the depth dependence of 
$\nabla =(d\,logT)/(d\,logP_g)$ (Fig. 7). For the umbral M3 model the 
$\nabla$-maximum occurs as deep as $\tau_{0.5}^{umbra}\approx10.0$ but for 
the photosphere much higher at $\tau_{0.5}^{phot}\approx2.0$. On the other 
hand, the criterion for convective instability, $\nabla=\nabla_{ad}=0.4$ is 
satisfied for M3 at $\tau_{0.5}^{umbra}=2.4$ and for the photosphere at 
$\tau_{0.5}^{phot}= 0.44$ (see Fig.\,6). 

Both considerations indicate that convection is possible for umbrae only at 
considerably larger depths than for the photosphere, in full agreement with 
the absence of macro-turbulence in line profile representations with the 
new empirical working model M3 for sunspot umbrae.

%
%

\begin{acknowledgements}
We are very grateful to E.\,H.\,Schr\"oter and H.\,Schleicher for interesting discussions. 
\end{acknowledgements}
%
%

\begin{figure*}[ht]     
   \hspace{3mm}\includegraphics[width=16.0cm]{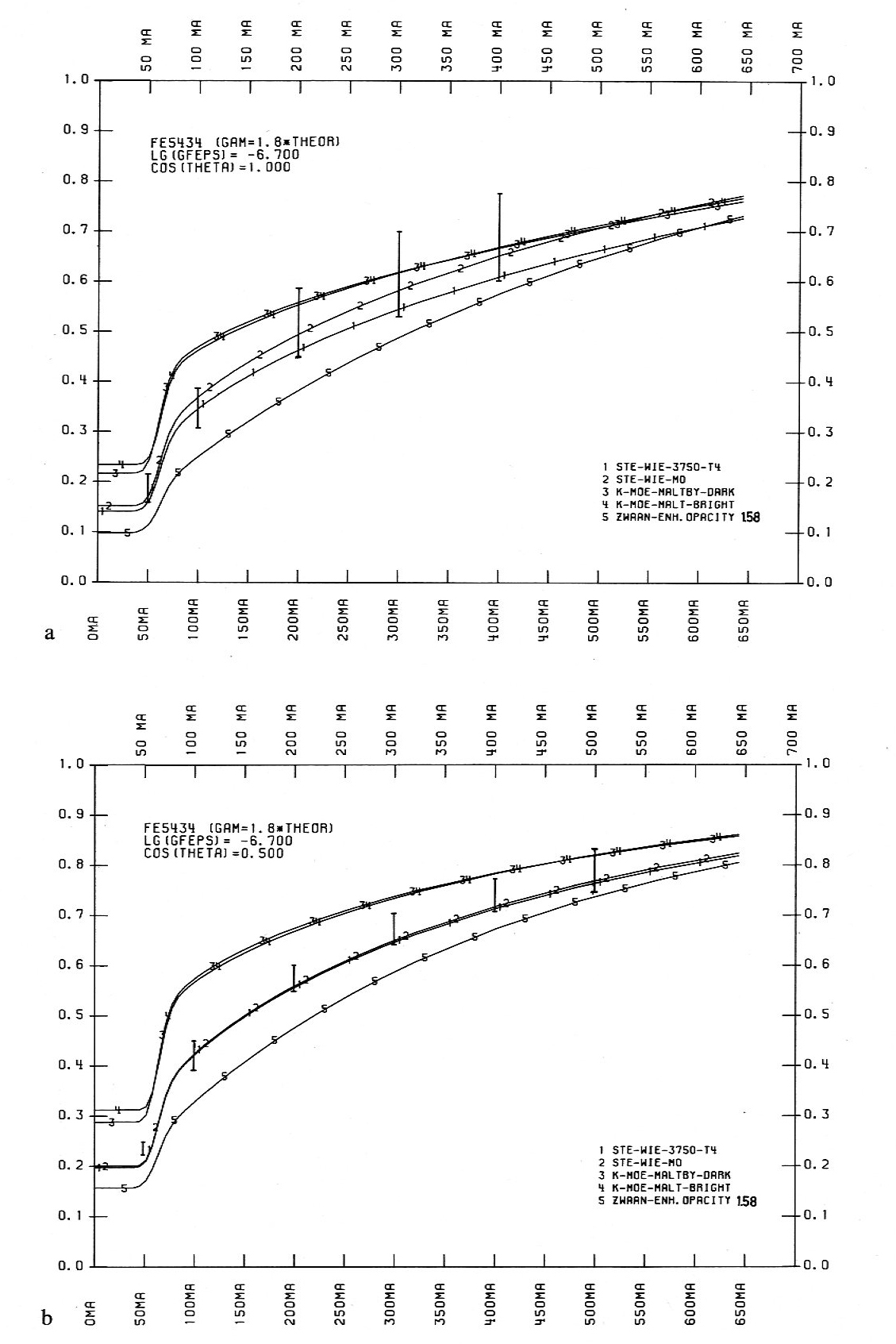}
   \caption{Fe\,5434 at $cos\vartheta=1.0$ ({\it upper}) and 
       $cos\vartheta=0.5$ ({\it lower panel}); 
       calculated ({\it solid lines}); observations ({\it bars}).}
   \label{Fig7}
    \end{figure*}

%

%

   \begin{figure*}[ht]     
   \hspace{3mm}\includegraphics[width=16.0cm]{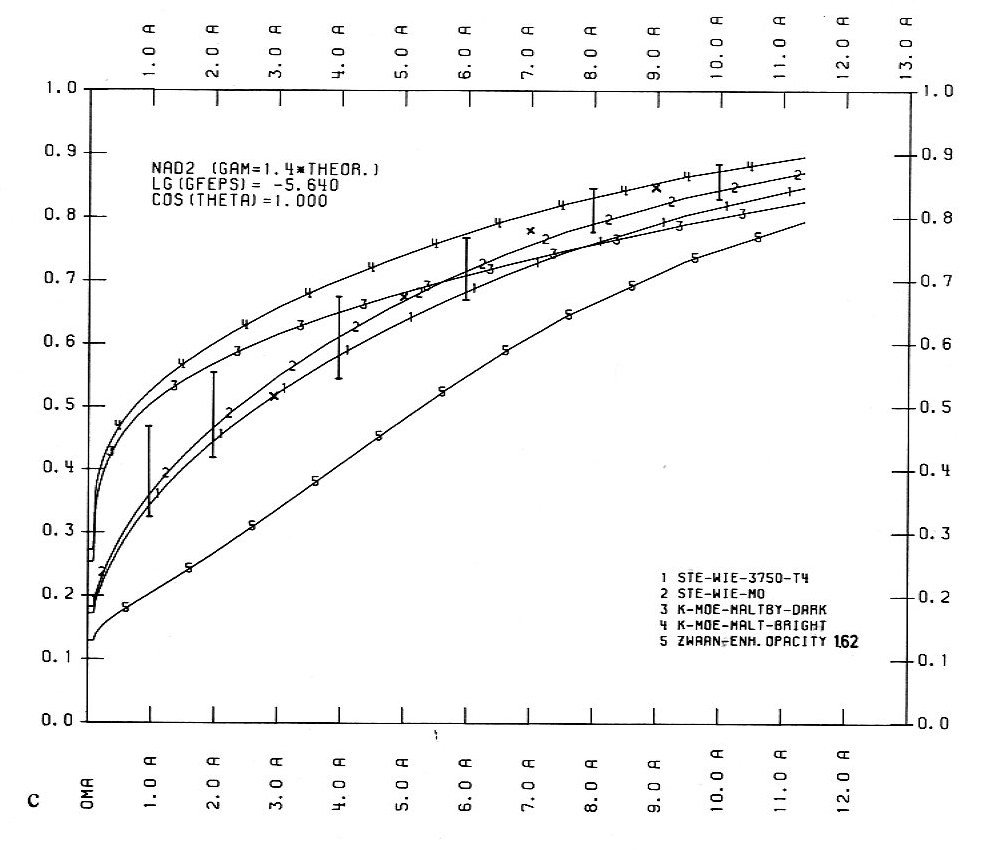}
   \caption{Calculated and observed profiles of the violet wing 
   of NaD2; x = raw data by J.\,Harvey (priv.\,com.)}
   \label{Fig8}
    \end{figure*}

%

%

   \begin{table*}[h]
   \caption[]{Umbral C\,I lines from observations ({\it upper}) and 
      from various emirical models ({\it lower part}).}
   \hspace{15mm}\includegraphics[width=13.0cm]{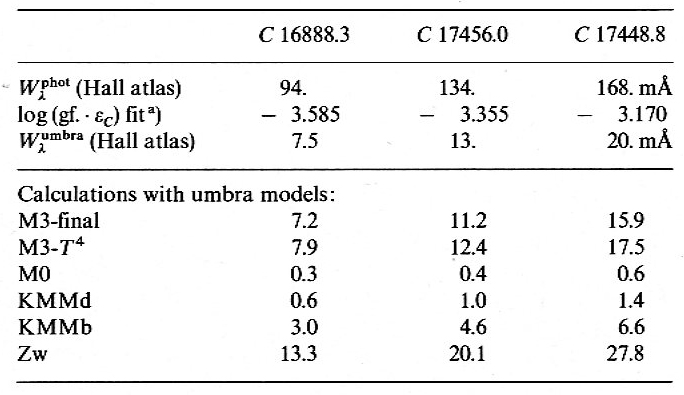}
   \label{Tab2}   
   \end{table*}

%

%

   \begin{table*}[h]
   \caption[]{Model M3 with deep layers as $T^4$
   stratification ({\it upper}) and fit to the Meyer et al. model for
   $\tau>1.0$ ({\it lower part}); also given are the abundances used.}
   \hspace{-5mm}\includegraphics[width=19.5cm]{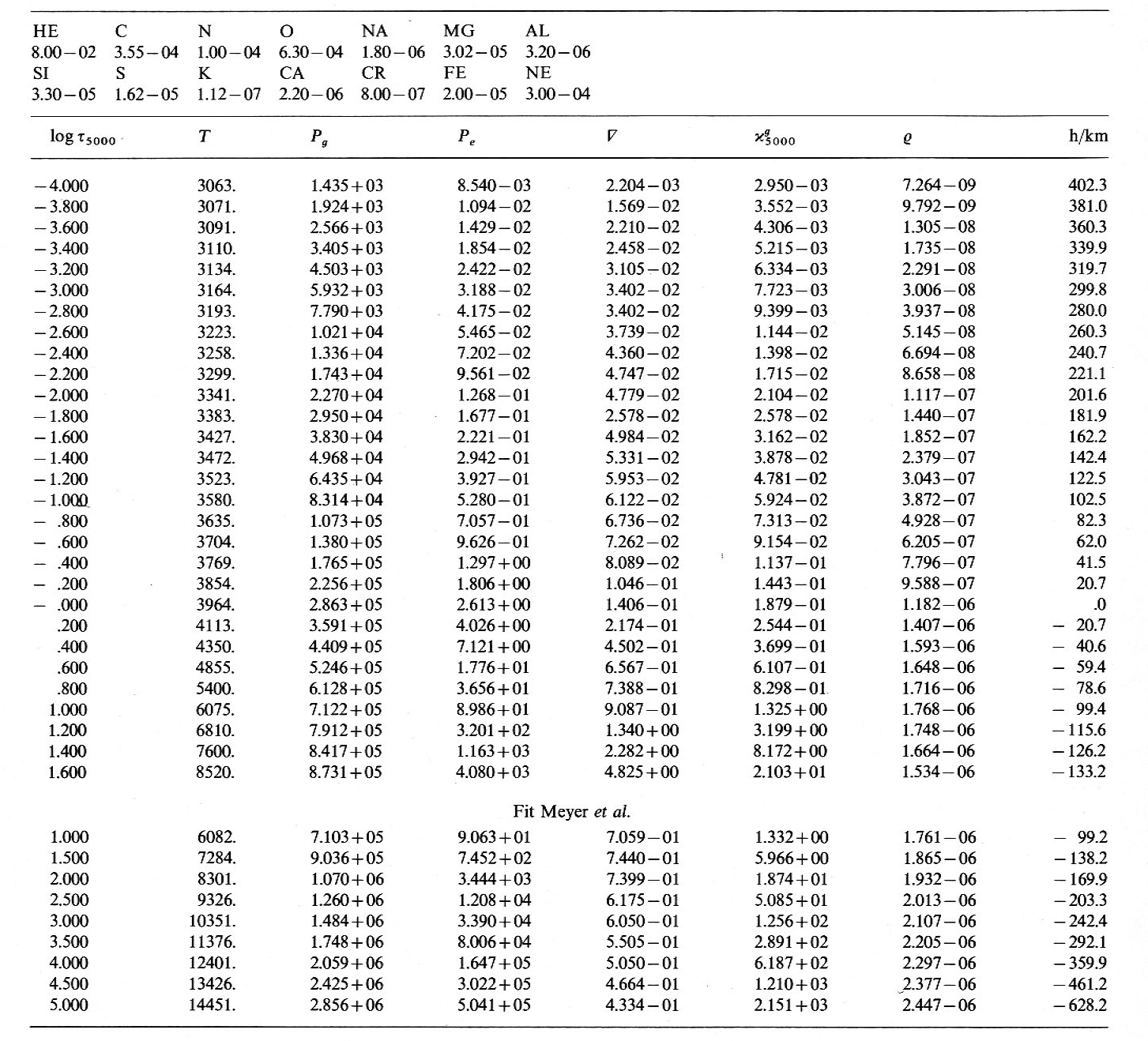}
   \label{Tab3}
    \end{table*}

%

%
%

\end{document}